\documentclass[aps, pra, reprint,superscriptaddress,nobibnotes]{revtex4-2}

\usepackage{amsthm, amsfonts, amsmath, amssymb}
\usepackage{mathtools, physics, bm}
\usepackage{hyperref}
\usepackage{xcolor, colortbl}
\usepackage{booktabs, longtable}
\usepackage{pifont}

\newcommand{\cmark}{\textcolor{green}{\ding{51}}}
\newcommand{\xmark}{\textcolor{red}{\ding{55}}}
\newcommand{\1}{\text{\usefont{U}{bbold}{m}{n}1}}

\begin{document}

\title{Comment on \lq A General Framework for Constructing Local Hidden-state Models to Determine the Steerability\rq }

\author{Nick von Selzam}
\affiliation{Max Planck Institute for the Science of Light, Erlangen 91058, Germany}
\affiliation{Department of Physics, Friedrich-Alexander-Universität Erlangen-Nürnberg, Erlangen 91058, Germany}

\author{Florian Marquardt}
\affiliation{Max Planck Institute for the Science of Light, Erlangen 91058, Germany}
\affiliation{Department of Physics, Friedrich-Alexander-Universität Erlangen-Nürnberg, Erlangen 91058, Germany}


\begin{abstract}

We point out that the method presented in a recent arXiv article by Jia et al. (arXiv:2512.21848) for constructing local hidden-state models closely follows the framework we developed in N.~von Selzam \& F.~Marquardt (PRX Quantum, 2025) for constructing local hidden-variable models.
While Jia et al. cite our work, the extent of the methodological overlap and the degree of textual similarity are not adequately reflected by the attribution given. We document this overlap in detail.

\end{abstract}

\maketitle

In our paper~\cite{vonSelzam2025}, we introduced a general machine-learning-inspired framework for constructing local hidden-variable (LHV) models for continuous sets of measurements, such as all projective measurements. The method was published in PRX Quantum on April~$24^{\text{th}}$, $2025$. The main technical contribution of our work was to construct and parameterize general LHV models for projective measurements in a way that allows for efficient numerical optimization over the space of all LHV models.

In Ref.~\cite{Jia2025}, posted on arXiv on December~$26^{\text{th}}$, $2025$, the authors present what they describe as a ``machine learning-based framework that employs batch sampling of measurements and gradient-based optimization to construct an optimal LHS model'' for the related problem of quantum steering. Their paper cites our work~\cite{vonSelzam2025} as ``Ref.~[17]'' in three places: once in the introduction as related work, and in Sec.~III.B as well as the conclusion crediting us for the idea of an expansion of the response function into polynomial basis functions. However, the methodological overlap extends significantly beyond this single idea.

Table~\ref{tab:method} lists the relevant methodological concepts, their location in both papers, and indicates whether or not attribution is given. This mostly concerns Sections~III and~IV in Ref.~\cite{Jia2025}. The corresponding content in our paper~\cite{vonSelzam2025} appears in Sections~III and~IV as well as Appendix~G.1.

Beyond the methodological overlap, many passages in Ref.~\cite{Jia2025} closely paraphrase or nearly reproduce text from our paper~\cite{vonSelzam2025}. We list examples in Table~\ref{tab:text} and Table~\ref{tab:text2}. In many cases the wording differs at most by a small change of notation (e.g., LHV$\to$LHS or renaming of variables such as $k\to a$) or minor synonym substitutions.

We note that applying our framework to the steering problem is a reasonable and interesting extension. The new element in Ref.~\cite{Jia2025} is the parametrization of the hidden states via unconstrained complex matrices, which is specific to the steering setting. Our concern is not with the extension itself, but with the presentation. Many text passages essentially copy our formulations and the presentation conveys the false impression that the underlying framework is largely an original contribution of Ref.~\cite{Jia2025}.

\begin{table*}[t]
\centering
\caption{Methodological elements appearing in both papers. \lq Cited\rq\ indicates whether the element is attributed to Ref.~\cite{vonSelzam2025} (ours) at the point where it appears in Ref.~\cite{Jia2025}. Ordered by appearance in Ref.~\cite{Jia2025}.}
\label{tab:method}
\setlength{\tabcolsep}{10pt}
\renewcommand{\arraystretch}{1.7}
\begin{tabular}{p{9cm} p{1.4cm} p{1.4cm} c}
\toprule
\textbf{Concept} & \textbf{Ref.~\cite{vonSelzam2025}} (ours) & \textbf{Ref.~\cite{Jia2025}} & \textbf{Cited} \\ \hline
\midrule
Representation of the hidden-variable distribution by a finite cloud of hidden-variable tuples & Sec.~III & Sec.~III & \xmark \\ \hline 
\addlinespace
Softmax parametrization of the local measurement rule & App.~G.1 & Sec.~III & \xmark \\ \hline
\addlinespace
Expansion of measurement operators into generalized Gell-Mann matrices, including a particular non-standard choice of normalization for~$G_0$ & App.~G.1 & Sec.~III.A & \xmark \\ \hline
\addlinespace
Expansion of the measurement rule into orthonormal basis functions on the space of measurements & App.~G.1 & Sec.~III.B & \cmark \\ \hline
\addlinespace
Hidden variables as coefficient matrix of the expansion & App.~G.1 & Sec.~III.B & \xmark \\ \hline
\addlinespace
For projective qubit measurements use odd spherical harmonics & Sec.~IV & Sec.~III.C & \xmark \\ \hline
\addlinespace
Use a sigmoid function to ensure non-vanishing gradients & Sec.~IV & Sec.~III.C & \xmark \\ \hline
\addlinespace
Mean distance between LHV and quantum statistics over all measurements as a loss function & Sec.~III & Sec.~IV.A & \xmark \\ \hline
\addlinespace
Stochastic gradient descent with batch-sampled measurements & Sec.~III & Sec.~IV.B & \xmark \\ \hline
\bottomrule
\end{tabular}
\end{table*}

\begin{table*}[t]
\centering
\caption{Side-by-side comparison of representative passages ordered by their appearance in our paper~\cite{vonSelzam2025}.}
\label{tab:text}
\setlength{\tabcolsep}{10pt}
\renewcommand{\arraystretch}{2}
\begin{tabular}{p{1.3cm} p{6cm} p{1.3cm} p{6cm}}
\toprule
\textbf{Ref.~\cite{vonSelzam2025}} & \textbf{Ref.~\cite{vonSelzam2025}} content (ours) & \textbf{Ref.~\cite{Jia2025}} & \textbf{Ref.~\cite{Jia2025}} content \\ \hline
\midrule

Sec.~III, first paragraph
& ``[\ldots] any distance measure ($D\geq 0$ and $D=0$ if and only if the two distributions are identical) can be used as long as it is differentiable with respect to the individual probabilities.''
& Sec.~IV.A, after Eq.~(9)
& ``Any valid distance metric ($D_{meas}\geq 0$ with $D_{meas}=0$ if and only if the assemblages are identical) may be utilized, provided it is differentiable with respect to the individual components.''
\\ \hline
\addlinespace

Sec.~III, first paragraph
&``Mathematically, our goal is to minimize $D[P^{\mathrm{QM}}(\cdot|x),P^{\mathrm{LHV}}(\cdot|x)]$ simultaneously for all possible measurement settings~$x$. The free parameter for this optimization is the hidden-variable distribution~$p$, [\ldots]''
& Sec.~IV.A, after Eq.~(9)
& ``Mathematically, our objective is to minimize $D[\{\sigma^{LHS}_{a|x}\}_{a=1}^O,\{\sigma^{QM}_{a|x}\}_{a=1}^O]$ simultaneously across all possible measurement settings~$M_x$. The free parameters in this optimization are the hidden variables [\ldots]''
\\ \hline
\addlinespace

Sec.~III, around Eq.~(5)
& ``Borrowing machine learning terminology, this can be achieved by minimizing the scalar loss function $\mathcal{L}(\mathrm{LHV}\,||\,\mathrm{QM}) = \langle D[P^{\mathrm{QM}}(\cdot|x),P^{\mathrm{LHV}}(\cdot|x)]\rangle_x$.''
& Sec.~IV.A, around Eq.~(10)
& ``Drawing upon machine learning terminology, this minimization can be formulated as optimizing a scalar loss function: $\mathcal{L}(\mathrm{LHS}\,||\,\mathrm{QM}) = \langle D[\{\sigma^{LHS}_{a|x}\}_{a=1}^O,\{\sigma^{QM}_{a|x}\}_{a=1}^O]\rangle_{M_x}$.''
\\ \hline 
\addlinespace

Sec.~III, after Eq.~(5)
& ``This is a single non-negative number: the mean deviation of $P^{\mathrm{LHV}}(\cdot|x)$ from $P^{\mathrm{QM}}(\cdot|x)$ over all measurements~$x$. In this way, we treat all measurement choices as equally important.''
& Sec.~IV.A, after Eq.~(10)
& ``This is a single non-negative number: the mean deviation of $\{\sigma^{LHS}_{a|x}\}^O_{a=1}$ from $\{\sigma^{QM}_{a|x}\}^O_{a=1}$ over all possible measurement settings~$M_x$. In this way, we treat all measurement choices as equally important.''
\\ \hline
\addlinespace

Sec.~III, after Eq.~(5)
& ``A vanishing loss $\mathcal{L}=0$ means that the LHV model reproduces the quantum-mechanical measurement statistics perfectly for all measurements.''
& Sec.~IV.A, after Eq.~(10)
& ``A loss function $\mathcal{L}=0$ indicates that the quantum assemblages obtained under all possible measurements can be reproduced by the LHS model.''
\\ \hline 
\addlinespace

Sec.~III, around Eq.~(6)
& ``After random initialization of these parameters they are adjusted iteratively as $\theta^{\mathrm{new}} = \theta^{\mathrm{old}} - \eta \nabla_\theta \mathcal{L}(\theta^{\mathrm{old}})$.''
& Sec.~IV.B, around Eq.~(11)
& ``After random initialization of these parameters, they are adjusted iteratively. [\ldots] The parameters are then updated according to $\boldsymbol{\theta}_{\mathrm{new}} = \boldsymbol{\theta}_{\mathrm{old}} - \eta \nabla_{\boldsymbol{\theta}} \mathcal{L}(\boldsymbol{\theta}_{\mathrm{old}})$, [\ldots]''
\\ \hline 
\addlinespace

Sec.~III, after Eq.~(6)
& ``For a sufficiently small learning rate $\eta>0$, this reduces the loss, as can be seen from a first-order expansion of $\mathcal{L}(\theta^{\mathrm{new}})$ in~$\eta$.''
& Sec.~IV.B, after Eq.~(11)
& ``[\ldots] which, for a sufficiently small learning rate $\eta>0$, reduces the loss (as can be verified by a first-order expansion of $\mathcal{L}(\boldsymbol{\theta}_\mathrm{new})$ with respect to~$\eta$).''
\\ \hline 
\addlinespace

Sec.~III, after Eq.~(6)
& ``The algorithm becomes stochastic in the following sense. We cannot calculate the mean over the full continuum of all measurement settings in the loss function exactly. Instead, in each update step we sample a new large but finite batch of measurement settings $(x^{(1)},\ldots,x^{(N_m)})$ and approximate the mean deviation and its gradient on this batch. In physicist terms, the mean in the loss function
is replaced by a Monte Carlo estimate.''
& Sec.~IV.B, last paragraph
& ``The algorithm becomes stochastic because we cannot compute the exact average over the full continuum of measurement settings in the loss function. Instead, at each update step, we sample a new, large but finite batch of settings $(M_1,M_2,\ldots,M_{N_{meas}})$ and replace the exact mean with a Monte Carlo estimate to approximate both the average deviation and its gradient.''
\\ \hline 

\bottomrule
\end{tabular}
\end{table*}

\begin{table*}[t]
\centering
\caption{Side-by-side comparison of representative passages ordered by their appearance in our paper~\cite{vonSelzam2025} (continued).}
\label{tab:text2}
\setlength{\tabcolsep}{10pt}
\renewcommand{\arraystretch}{2}
\begin{tabular}{p{1.3cm} p{6cm} p{1.3cm} p{6cm}}
\toprule
\textbf{Ref.~\cite{vonSelzam2025}} & \textbf{Ref.~\cite{vonSelzam2025}} content (ours) & \textbf{Ref.~\cite{Jia2025}} & \textbf{Ref.~\cite{Jia2025}} content \\ \hline
\midrule

Sec.~IV, around Eq.~(11)
& ``First, we introduce a cutoff in the spherical harmonics expansion, writing $\vec S_D(\hat n)$ for a vector containing all odd spherical harmonics up to degree $D$. For example (up to normalization factors), $\vec S_3(\hat n) \sim (x, y, z, xyz, \ldots, 3x^2y-y^3),\ (x, y, z)=\hat n$. This leads to finite dimensional hidden variables $\vec \lambda \in \mathbb{R}^d$, with $d = \frac{1}{2}(D+1)(D+2)$.''
& Sec.~III.C, around Eq.~(7)
& ``By introducing a order $D$, we denote the vector composed of odd spherical harmonics up to order $D$ as $\vec B^D(\vec g^0)$ (specific form for $D=3$ is given below): $\vec B_3(\vec g^0) \sim (x, y, z, xyz, \ldots, 3x^2y-y^3),\ \vec g^0 = (x, y, z)$. When considering the measurement rule for a single outcome, the hidden variable reduces to a vector $\vec \lambda \in \mathbb{R}^N$ (instead of a matrix), whose dimension is given by: $N = \frac{1}{2}(D+1)(D+2)$.''
\\ \hline
\addlinespace

Sec.~IV, before Eq.~(12)
& ``Second, for the numerical optimization, we require nonvanishing gradients with respect to the hidden-variable vectors. To achieve this, we replace the Heaviside function by a sigmoid function, bringing us back to probabilistic measurement rules: $q(\hat n, \vec \lambda) = \sigma[\vec S_D(\hat n)\cdot \vec \lambda],\ \sigma(x) = \frac{1}{1+e^{-x}}$.''
& Sec.~III.C, last paragraph
& ``To satisfy the requirement of non-vanishing gradients in numerical optimization, we employ a sigmoid function to transform the mapping into probabilistic form: $p(0|x,\lambda) = \text{sigmoid}\left[ \sum_{m=1}^N \vec B_m^D(\vec g^0)\cdot \vec \lambda \right],\ \text{sigmoid}[y] = \frac{1}{1+e^{-y}}$.''
\\ \hline 
\addlinespace

Sec.~V, after Eq.~(14)
& ``The parameter $v\in[0,1]$ is the visibility of the singlet. It is known that these states are separable for $v\leq\frac{1}{3}$ [\ldots]''
& Sec.~V.A, after Eq.~(12)
& ``The parameter $v\in[0,1]$ represents the visibility of the singlet. It is known that these states are separable for $v\leq 1/3$ [\ldots]''
\\ \hline
\addlinespace

App.~G.1, after Eq.~(G2)
& ``The functions $q_k$ are constrained by the normalization condition. [\ldots] it is useful to express the probability distributions~$\vec{q}$ via unconstrained functions $\vec{f}: X \times \Lambda \to \mathbb{R}^\Delta$: $q_k(x,\lambda)=\mathrm{softmax}[\vec{f}(x,\lambda)]_k = \frac{e^{f_k(x, \lambda)}}{\sum_{k'=1}^\Delta e^{f_{k'}(x, \lambda)}},$ [\ldots]''
& Sec.~III, right before Sec.~III.A
& ``To ensure that the response function $p(a|x,\lambda)$ satisfies the probability constraints $0\leq p(a|x,\lambda)\leq 1$ and $\sum_a p(a|x,\lambda)=1$, we define it using a softmax form: $p(a|x,\lambda) = \mathrm{softmax}[f_a(x,\lambda)] = \frac{e^{f_a(x, \lambda)}}{\sum_{a'=1}^O e^{f_{a'}(x, \lambda)}},$ [\ldots]''
\\ \hline 
\addlinespace

App.~G.1, around Eq.~(G8) and Eq.~(G9)
& ``[\ldots] the measurement operators $x_i$ are Hermitian matrices. Hence we can expand them into the generalized traceless Gell-Mann matrices $G_k$ and the identity: for any Hermitian $d$-by-$d$ matrix~$a$, there exists a unique $\vec{g}\in\mathbb{R}^{d^2}$ such that $a = a(\vec g) \equiv \frac{1}{\sqrt{d}}\sum_{\mu=0}^{d^2-1} g_\mu G_\mu$. Here we set $G_0 = \sqrt{(2/d)}\1$ such that $\Tr(G_\mu G_\nu) = 2\delta_{\mu\nu}$ for all $\mu,\nu \in \{0,\ldots,d^2-1\}.$''
& Sec.~III.A, around Eq.~(3) and Eq.~(4)
& ``These operators are Hermitian operators of dimension~$d$, and thus can be expanded in terms of the generalized traceless Gell-Mann matrices $G_\mu$ and the identity matrix~$I$: for any Hermitian $d\times d$ matrix $M_{a|x}$, there exists a unique vector $\vec{g}\in\mathbb{R}^{d^2}$ such that: $M_{a|x} = M_{a|x}(\vec g) \equiv \frac{1}{\sqrt{d}}\sum_{\mu=0}^{d^2-1} g_\mu G_\mu$. Here we set $G_0 = \sqrt{\frac{2}{d}}I$ so that $\tr(G_\mu G_\nu) = 2\delta_{\mu\nu}$, for all $\mu,\nu \in \{0,1,2,\ldots,d^2-1\}.$''
\\ \hline

\bottomrule

\end{tabular}
\end{table*}

\bibliography{bibliography}

\end{document}